# Chiral Quantum well Rashba splitting in Sb monolayer on Au(111)

Jinbang Hu[1,+], Lina Liu[2], Xiansi Wang[3], Yong P. Chen[2], Justin W Wells[1,4]

[1]*Department of Physics, NTNU, Trondheim, Norway.*
[2]*Institute of Physics and Astronomy and Villum Centers for Hybrid Quantum Materials and Devices, Aarhus University, 8000 Aarhus-C, Denmark.*
[3]*Hunan University, Changsha, 410082, China.*
[4]*Semiconductor Physics, Department of Physics, University of Oslo (UiO), NO-0371 Oslo, Norway.*

[+]Corresponding author: Jinbang Hu@ntnu.no.

**Abstract**
We present a comprehensive investigation into the atomic and electronic structures of a single-layer Sb(110) rhombohedral crystal formed on an Au(111) substrate. Low-energy electron diffraction (LEED) and scanning tunneling microscopy (STM) reveal a pure two-dimensional (2D) Sb stripe structure, composed of a pair of Sb(110) unit cells located in a chiral configuration with mirror symmetry breaking perpendicular to the direction of the bright stripe. Based on angle-resolved photoemission spectroscopy (ARPES) measurements and Sb-weighted band structure from density functional theory calculations, we report the unambiguous determination of Rashba spin-orbit coupled bands from the 2D Sb film, exhibiting a chiral symmetry in the electronic structure with the crossing points located at the $\overline{\Gamma}$ point and the $\overline{X}$ point, respectively. Moreover, From $dI/dV$ spectra and density of states (DOS) calculations, the quantum well (QW) Rashba-type states induced by the in-plane mirror symmetry breaking in the Sb stripe structure have been identified. Orbital decomposition of the projected band structure reveals that hybridization between Sb $p_y$ states and Au states modifies the spin splitting of the QW states, attributed to the intrinsic stong SOC of Au states introduced into the QW states.

**INTRODUCTION**

Manipulation of the spin degrees of the 2D gas confined in the ultrathin film plays an important role in the rapidly developing field of spintronics[1-2]. As predicted by the theory of the Rashba effect, the lifting of spin degeneracy originates from spin-orbit coupling (SOC) in inversion symmetry-broken structures[3]. Especially for the structural inversion asymmetry induced by the surface-potential gradient perpendicular to the substrate, a class of 2D surface alloys has been successfully fabricated on noble metal surfaces by substituting one out of three atoms within the topmost (111) surfaces, forming a $\sqrt{3} \times \sqrt{3}$ periodicity. Examples include Pb, Bi, and Sb grown on Ag(111) and Cu(111), and Sn on Au(111)[4-10]. The largest known Rashba spin splitting was reported for Ag$_2$Bi[6], and numerous prior studies of the 2D surface alloys have confirmed the important role played by atomic SOC in determining the strength of the splitting. However, the electronic properties of 2D adlayers, so far, have rarely been well discussed with in-plane asymmetry being considered. Similarly, like the Rashba splitting observed in 2D surface alloys, the ultrathin adlayer on the substrate can also exhibit spin-orbit splitting of the electronic states. Intriguingly, the asymmetric confinement from the interface potential can further modify electronic states, such as the giant spin-orbit splitting of QW states observed in the surface state of a Bi monolayer on Cu(111) due to the in-plane mirror symmetry breaking in the Bi adlayer [11].

Apart from the easy preparation of a long-range ordered surface alloy, the single-layer (SL) material may form several different reconstructions, and also the rotation of the domain determined by the symmetry of the substrate, leading to an extremely complicated 2D surface electronic structure. In the present work, based on STM observation of 2D Sb stripe structure formed on Au(111), we properly disentangle the complicated surface electronic structure and study the electronic structure originating from the Sb(110) rhombohedral phase. Taking advantage of DFT calculations, the important role of symmetry breaking, combined with atomic orbital hybridization, in determining the Rashba splitting with a chiral character has been confirmed. Furthermore, a direct comparison between STS measurements and integrated DOS around the $\bar{\Gamma}$ point helps to identify the Rashba split bands as QW states.

**EXPERIMENTAL AND THEORETICAL DETAILS**

All sample preparation steps and experiments were performed under ultra-high vacuum conditions. The Au(111) surface was prepared by repeated sputtering and annealing cycles. The quality of the clean surface was confirmed by the STM observation of the well-known Au-herringbone reconstruction (Figure S1). Subsequently, Sb (purity 99.9999%) was deposited onto the clean surface at room temperature, yielding a 2D Sb(110) rhombohedral phase by subsequent annealing at 300 °C. During the evaporation, the pressure remained better than 4×10$^{-10}$ mbar. The success of the sample preparation was confirmed by LEED and STM measurements. LT-STM/STS measurements carried out at 4 K were conducted using a Unisoku 1300 system. STS spectra were acquired using a commercial Pt-Ir tip with a standard lock-in technique with a frequency of 917 Hz. Band structure measurements were performed at $T \approx 115$ K using an aberration-corrected, energy-filtered photoemission electron microscope (EF-PEEM) (NanoESCA III, Scienta Omicron GmbH) equipped with a focused helium discharge lamp primarily generating He I photons at $h\nu = 21.22$ eV. A pass energy of $E_P = 25$ eV and a 0.5 mm entrance slit to the energy filter were used, yielding nominal energy and momentum resolutions of $\Delta E = 50$ meV and $\Delta k = 0.02$ Å$^{-1}$.

The density functional theory calculations were performed using the Vienna Ab initio Simulation Package (VASP) [13]. The interactions between the valence electrons and ion cores were described by the projector augmented wave method [14]. The electron exchange and correlation energy were treated by the generalized gradient approximation with the Perdew-Burke-Ernzerhof functional [15-16]. The kinetic energy cutoff of the plane-wave basis was set to 500 eV by default. For the matched models, the first Brillouin zone of the 2D Sb(110) on the Au(111) system was sampled with the $\bar{\Gamma}$-centered 6×6×1 k points. The structures were optimized with symmetry (space group PM) until the forces on the atoms were less than 10 meVÅ$^{-1}$. The Au(111) surface was described by a periodic slab separated by 15 Å vacuum. An 15 layers of Au atoms were included with the bottom 12 layers fixed as the bulk crystal structure while the top three layers were relaxed. One monolayer of Sb atoms was adsorbed on the top side of the Au slab. In order to "unfold" the band structure with the symmetry of the primitive unit cell of 2D Sb(110) and also the Au(111), we calculated the effective band structure proposed by Popescu and Zunger [17] as implemented in the BandUp code [18-19]. The DOS is integrated over the zone center near the $\bar{\Gamma}$ point with a momentum radius of $k = 0.185$ Å$^{-1}$. Spin-orbit coupling was included for all band structure calculations. The STM image was simulated using the Tersoff-Hamann approach [20].

**RESULTS AND DISCUSSION**

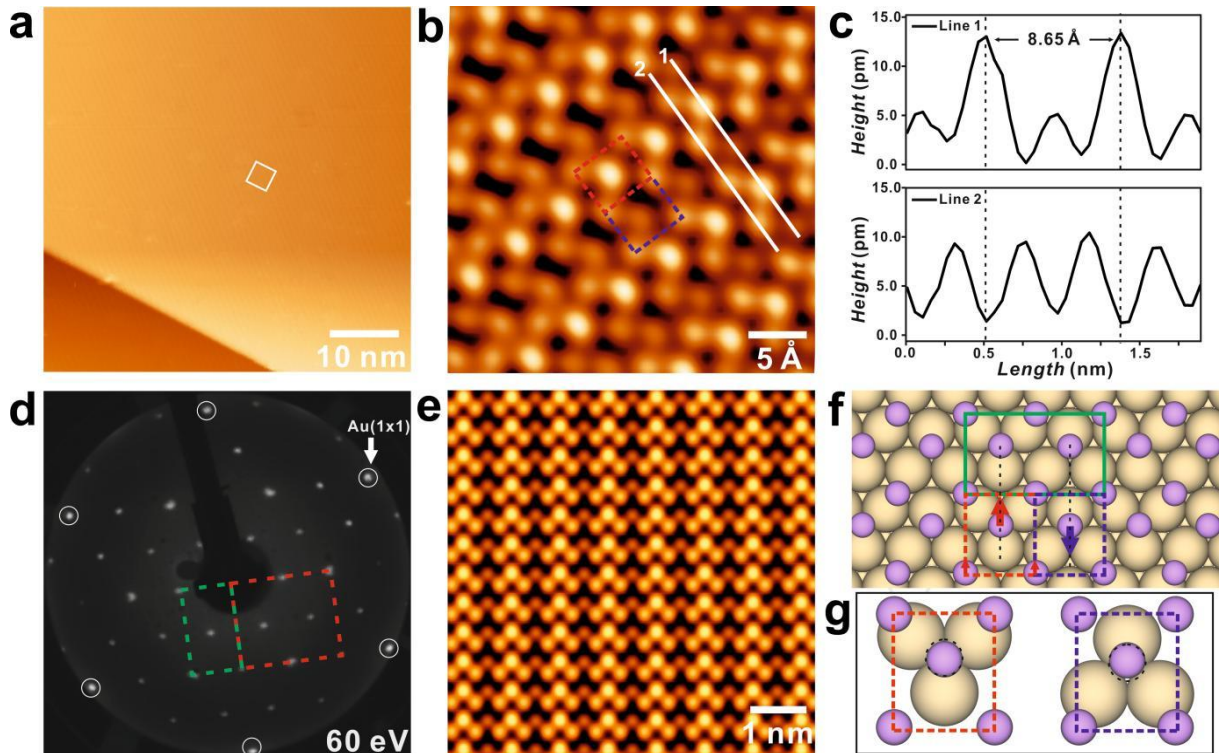

Figure 1. (a) Large-scale STM image of monolayer Sb on the Au(111). (b) High-resolution STM image depicted by the white square in panel (a), demonstrating a well-ordered $3 \times \sqrt{3}$ reconstruction composed of two adjacent Sb(110) unit cells marked by the blue and red dashed rectangular. (c) Line profile corresponding to the white line in panel (b), revealing the periodicity 8.65 Å and mirror symmetry of the $3 \times \sqrt{3}$ reconstruction. (d) LEED pattern of the 2D Sb film grown on the Au(111) taken at a beam energy of 60 eV. The corresponding reciprocal unit cells of $3 \times \sqrt{3}$ reconstruction and the Sb(110) unit cell in (b) are indicated by the green and red dashed rectangular, respectively. (e) STM simulation of monolayer Sb on Au(111). (f) Fully optimized slab lattice of the Sb $3 \times \sqrt{3}$ reconstruction, with the blue and red arrows overlaid on two Sb atoms in two adjacent Sb(110) unit cells to explain the potential they experienced. The mirror plane perpendicular to $x$ axis is marked by the black dashed line. (g) Schematic illustration of a pair of antiparallel potential gradients generated by the misalignment of the Sb atom in the center of the Sb(110) unit cell. The Sb located at the threefold hollow sites of the substrate is marked by a black dashed circle. Scanning parameters: (a) 1.4 V, 0.78 nA; (b) 1.0 V, 0.9 nA.

figure 1 summarizes the structure of monolayer Sb on Au(111) fabricated by annealing of a thicker Sb film sample at 300 °C. The corresponding LEED pattern (Figure 1d) and STM image (Figure 1a,b) are much like what has been observed previously under a similar sample preparation conditions[12]. In Figure 1a, our STM observation reveals a typical large-scale STM image of the sample, demonstrating a homogeneous stripe structure on the entire Au(111) surface. Figure 1b shows a zoomed-in STM image of the area designated by the white square in Figure 1a, revealing a well-ordered $3 \times \sqrt{3}$ reconstruction of the Sb stripe structure formed on Au(111). Moreover, the $3 \times \sqrt{3}$ reconstruction can be divided into two adjacent Sb(110) subunit cells, marked by the red and blue rectangular, differing obviously in the respective heights of the Sb atom in the center. Correspondingly, the new spots appearing in the LEED pattern originate from the reciprocal unit cell of the $3 \times \sqrt{3}$ reconstruction (green dashed rectangular) and the Sb(110) subunit cell (red dashed rectangular) with threefold rotational symmetry.

In our study, we report a rectangular supercell of the 2D Sb stripe structure with a mirror symmetry. The line profile marked in Figure 1b shows the differing heights of the Sb atom in the Sb monolayer structure, as shown in Figure 1c, revealing the mirror symmetry in the Sb $3 \times \sqrt{3}$ reconstruction and also in the two Sb(110) subunit cells. The mirror plane for the $3 \times \sqrt{3}$ reconstruction is along the brightest (or dimmest) line of Sb atoms, which is also the mirror plane for the two Sb(110) subunit cells, respectively. Based on the STM observation, the theoretically optimized structure of 2D Sb on Au(111) is depicted in Figure 1f, and the corresponding STM simulated image (Figure 1e) well reproduces the stripe feature with mirror symmetry in 2D Sb on Au(111). On closer inspection of the optimized structure, the commensurate rectangular $3\times\sqrt{3}$ stripe superstructure can also be expressed in its matrix form as $\begin{pmatrix} 3 & 0 \\ 1 & 2 \end{pmatrix}$ referring to the Au(111) surface or $\begin{pmatrix} 2 & 0 \\ 0 & 1 \end{pmatrix}$ referring to the Sb(110) rectangle unit cell. The 2D Sb stripe structure can be divided into two Sb(110) sub-unit cells with the central Sb atom within the rectangle slightly misaligned to the threefold hollow sites of the Au substrate, as shown in Figure 1g. Hence, It can be expected that the Sb atom, marked by the arrow, in the centre of two adjacent Sb(110) unit cells experiences an antiparallel potential gradient along the $y$ direction since each Sb(110) unit cell represents a mirror symmetry along the $y$ direction. The mirror plane is marked by the black dashed line in Figure 1f. From the perspective of the $3\times\sqrt{3}$ stripe structure, a pair of antiparallel potential gradients generated by the misalignment of the Sb atom in the two adjacent Sb(110) unit cells can be invoked to illustrate the presence of chiral symmetry. The manifestation of chiral symmetry is also evident in the electronic properties, which will be expounded upon in the subsequent discussion.

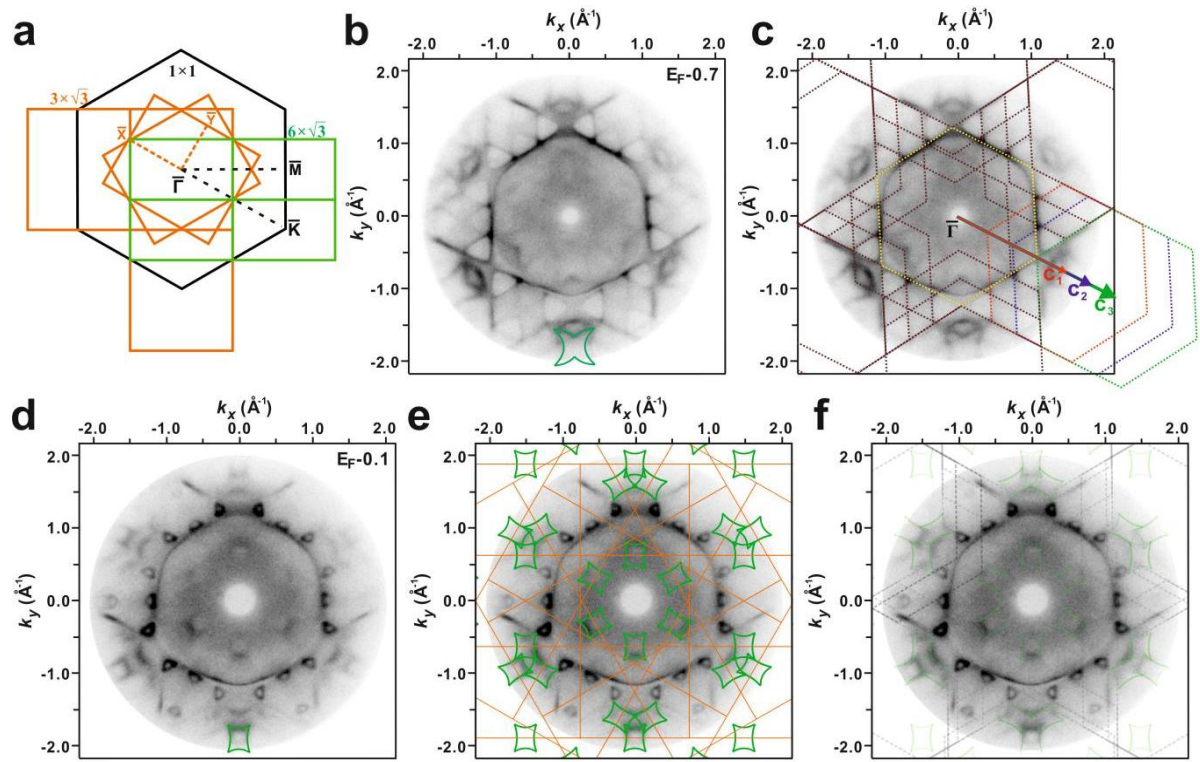

Figure 2. (a) $1\times 1$, Sb(110) and $3\times\sqrt{3}$ surface Brillouin zone (SBZ) scheme for the Au(111) surface, Sb(110) rhombohedral unit cell and Sb stripe structure, respectively. (b-f) Constant energy contours from the Sb/Au(111) $3\times\sqrt{3}$ surface at 115K. The contours are shown at 0.7 eV (b,c) and 0.1 eV (d-f) below the Fermi level. The newly appearing surface band feature is marked as a green flower. (c) A dashed hexagon (in brown) is drawn to trace the weak features appearing in the constant energy contours. The weak dashed hexagon originates from Umklapp scattering of the Au $sp$ band, marked by the bright dashed hexagon (in yellow). Three reciprocal lattice vectors marked as $C_1$, $C_2$ and $C_3$ indicate the possible Umklapp scattering process. (e) Sb(110) SBZ with the green-flower marks overlaid at $\bar{X}$ point of the Sb(110) SBZ to trace the band feature in the constant energy contours. (f) Umklapp scattering of the Au $sp$ band, combined with the green-flower marks from the three different azimuthal orientations of the Sb(110) rhombohedral phase fits on the constant energy contours.

We now turn to the electronic structure investigated by ARPES. Figure 2 shows the constant energy contours obtained from the Sb/Au(111) $3\times\sqrt{3}$ surface. In this section, we use a constant energy contour at 0.7 eV below Fermi level as an example to delve into a detailed electronic structure study of Umklapp scattering originating from the Au $sp$ band. Additionally, we employ a constant energy contour at 0.1 eV to elucidate the complex Umklapp scattering band induced by the three different azimuthal orientations of the Sb(110) rhombohedral lattice on the Au(111) surface. Figure 2a presents the $1 \times 1$ (black hexagon), Sb(110) (yellow rectangle) and $3 \times \sqrt{3}$ (green rectangle) SBZ schemes for the Au(111) surface, Sb(110) rhombohedral unit cell and Sb stripe structure, respectively. The three overlaid yellow rectangles indicates three rotational domains of the Sb(110) rhombohedral unit cell, extending across the 1st Au(111) SBZ. As a comparison of Figure 2b and 2d, the weak triangle-shaped feature appearing at around the $\overline{\text{K}}$ point of Au(111) SBZ shrinks to a smaller size with an increasing intensity of the signal as the binding energy approaches Fermi level. To understand the origin of the weak triangle features around the $\overline{\text{K}}$ point, the brown dashed hexagons were applied to trace the band features observed in the experimental constant energy contours. These brown dashed hexagons can be categorized into three groups, originating from three different Umklapp scattering processes of Au $sp$ band (yellow dashed hexagon) with reciprocal lattice vectors marked as $c_1$ (in red), $c_2$ (in blue) and $c_3$ (in green), respectively. A careful analysis reveals that $c_1 = 4a_1 - a_2, c_2 = 3a_2, c_3 = 4a_1$ . Notice, we only use three types of Umklapp scattering vectors to set an example of well-fitting the weak triangle band features in the experimental constant energy contours. There is a highly possible of extra Umklapp scattering vectors being applied to trace the triangle band feature in higher order SBZs of the Au(111) surface, as well as the fuzzy background in the first Au(111) SBZ.

Furthermore, the influence from the three rotational domains of the Sb(110) rhombohedral lattice has also been discussed to disentangle the flower-shaped band (marked by the green line in Figure 2b and 2d), which cannot be ascribed solely to the Umklapp scattering of the Au $sp$ band. Figure 2e represents the constant energy contour at the binding energy of -0.1 eV below Fermi level overlaid by the Sb(110) SBZ (orange rectangle) with the flower-shaped band located at the $\overline{\text{X}}$ point. The new schematic drawing of flower-shaped bands agree well with the experimental band features appearing on either side of the $\overline{\text{M}}$ points along the $\overline{\Gamma} - \overline{\text{M}}$ direction of the Au(111) SBZ. Ultimately, Figure 2f confirms the veracity of the novel schematic rendering of the flower-shaped bands, harmoniously aligned with the encompassing larger hexagon-traced Umklapp pattern, thus eloquently corroborating the nature of the complicated surface bands feature engendered by the Sb/Au(111) $3 \times \sqrt{3}$ surface. The well agreement of the Umklapp scattering band fitting with the complicated feature in ARPES data indicates the well-ordered Sb(110) rhombohedral structure formed on Au(111).

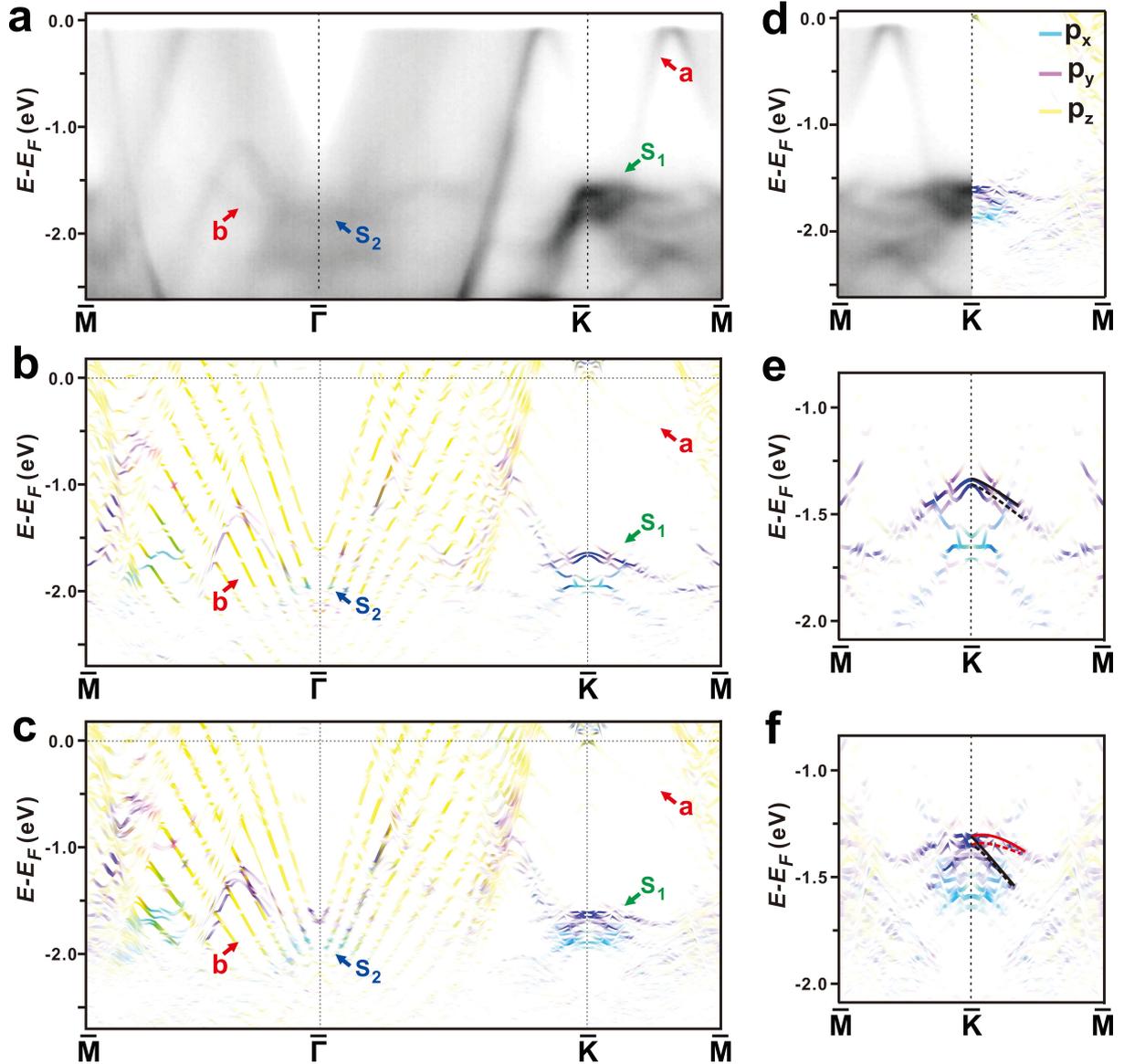

Figure 3. (a) ARPES results from the commensurate monolayer of Sb on Au(111) sliced along high-symmetry directions of the Au(111) SBZ. Dispersive features labeled b originate from three-dimensional bulk bands of Au(111). The labels $S_1$ and $S_2$ denote the bands originating from the 2D Sb film. The label a originates from Umklapp scattering of the Au $sp$ band. Calculated band structure without SOC (b) and with SOC (c) for the optimized model of the 2D Sb on Au(111). Blue, pink and yellow symbols show contributions from $p_x$, $p_y$, and $p_z$ states, respectively. (d) Comparison between ARPES data and calculated band dispersion along the $\overline{K}\,\overline{M}$ direction. The calculated bands are shifted up by 0.3 eV to match the ARPES data. (e-f) Comparison between calculated band dispersion along the $\overline{K}\,\overline{M}$ direction without SOC (e) and with SOC (f) around $\overline{K}$ point to illustrate the splitting of two pair of Rashba-type bands, marked by the dashed and solid lines in red (spin up) and black (spin down) coded.

To gain further insight into the electronic structure of the 2D Sb film and to elucidate the origin of the Rashba-type spin-orbit splitting observed in our ARPES measurement, band structure calculations were performed based on a Sb $3 \times \sqrt{3}$ superstructure possessing mirror symmetry of the two Sb(110) subunit cells as previously expounded. Figures 3b and 3c show such weighted band structure without SOC and with SOC projected from the Sb monolayer, respectively. The calculated bands with SOC, as displayed in Figure 3c, exhibit a good overall agreement with the experimental bands in Figure 3a, apart from the band feature 'a' and 'b', as labeled. The feature 'a' presents a discernibly akin band dispersion to that of the Au $sp$ band, and its conspicuous absence within the calculated band structure correlates with our ascertained deduction of its origin emanating from Umklapp scattering of the Au sp band, which is in harmony with our discussion of the weak triangle-shaped feature observed in constant energy contours in Figure 2. Moreover, the new band features around the $\bar{\Gamma}$ and $\bar{K}$ point, labeled as S$_1$ and S$_2$ respectively, appearing in our ARPES measurement and calculated band structure projected from states of Sb, indicate their contribution mainly originate from the topmost 2D Sb film. As shown in Figure 3a, the several branches of the S$_1$ band feature show a strong photon emission intensity near the $\bar{K}$ point and disperse downward away from a maximum energy at -1.55 eV below the Fermi level, while the band dispersion of feature S$_2$ looks blurry due to the overlapping with the replicas of the Au bulk band feature b, resulting from photo-excitation by the He I$_\beta$ emission line of the nonmonochromatized He I light source.

In comparison with the weighted band structure without SOC in Figure 2b, one notices that the band feature S$_1$ splits into two discernible parabolic sub-bands around the $\bar{K}$ point after SOC is included, as shown in Figure 2c. Moreover, a direct comparison in Figure 3d shows a good agreement between the DFT calculations and experimental results along $\bar{K}-\bar{M}$ direction, which confirms a typical feature of Rashba SOC present in the band feature S$_1$. Further analysis of the band structures projected on Sb $p$ orbital (Figure 3c) indicates that the Rashba-type band feature S$_1$ mainly consists of Sb $p_y$ orbitals, and the band feature S$_2$ donates from Sb $p_x$ orbital. The $p_z$ orbitals become indistinguishable and are strongly hybridized with the Au substrate. In Figure 3e and 3f, the zoomed-in band feature S$_1$ along $\bar{K}-\bar{M}$ direction shows two parallel branches, marked by dashed and solid lines, respectively, slightly gapped in energy around the $\bar{K}$ points without SOC included. With SOC, these two branches split into two pairs of Rashba-type bands with the separation in energy being further enhanced. We deduce the small energy discrepancy of the two pairs of Rashba-type bands is caused by the Sb atom in the middle of the two Sb(110) unit cells experiencing a different potential gradients from the Au substrate since the two Sb(110) unit cells in the $3\times\sqrt{3}$ stripe structure buckle oppositely normal to the Au(111) surface. Our further calculation of the Sb/Au(111) $3\times\sqrt{3}$ surface without buckling of the stripe structure confirmed the obviously decreasing separation of the two branches of the S$_1$ band (Figure S2).

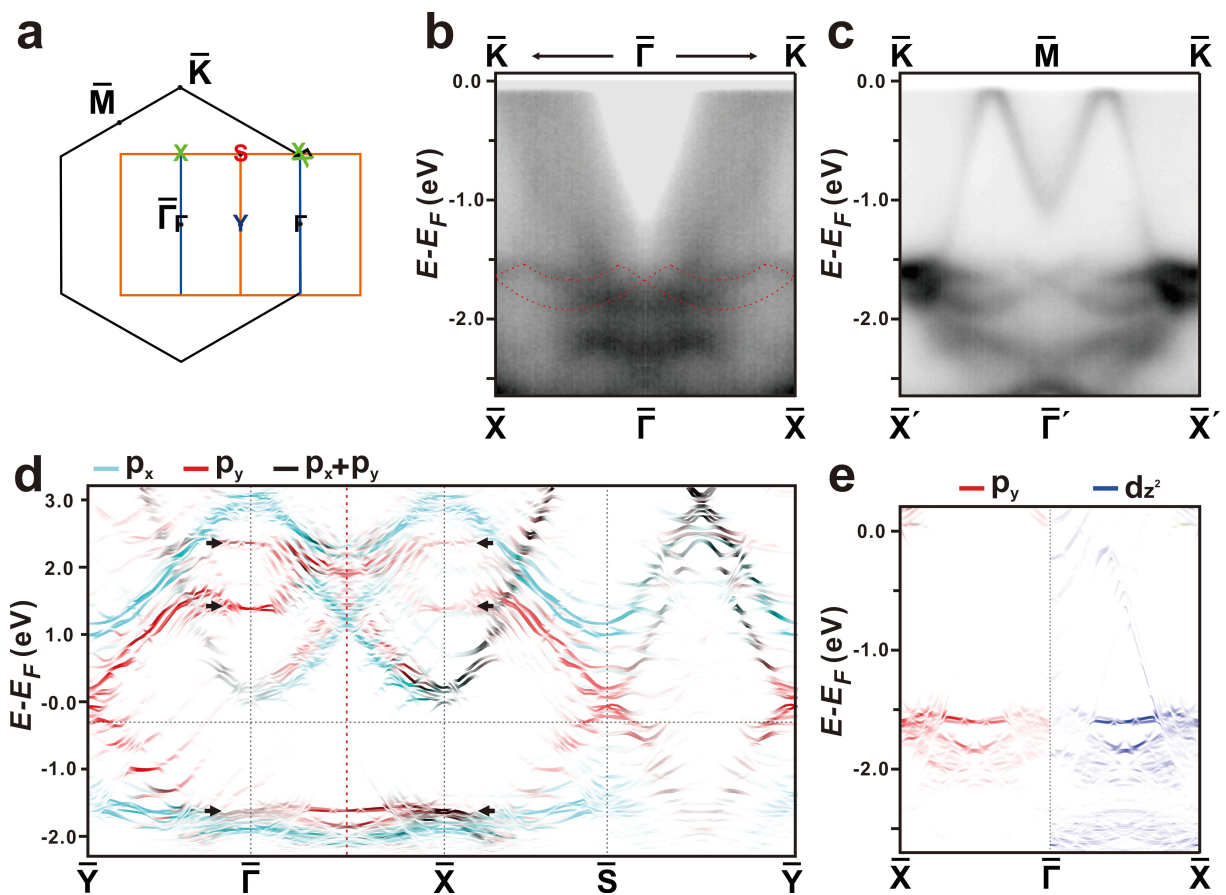

Figure 4. (a) The distribution of the SBZ of the 2D Sb(110) rhombohedral unit cell and Au(111) with the band structure around the high symmetry point of the Sb(110) SBZ symbolled by $\Gamma$, X, Y, and S respectively. (b-c) The ARPES results of Sb/Au(111) $3 \times \sqrt{3}$ surface sliced along the high-symmetry directions in the first Sb(110) SBZ (b) and the second SBZ (c) marked as blue line in (a). The red dashed curve in (b) is used to trace the Rashba splitting band discussed in Figure 3. (d) Calculated band structure unfolded to the SBZ of the Sb(110) sub-unit cell exhibits a character of chiral symmetry. Light blue and pink symbols show contributions from $p_x$ and $p_y$ states of the Sb topmost layer, respectively. The black arrows are used to mark the energy position of QW states. The red dashed line is used to mark the mirror plane. (e) Calculated band structure projected from the $d_{z^2}$ orbital (weighted by blue in the right side of the panel) of topmost Au layer and the $p_y$ orbital (weighted by red in the left side of the panel) of topmost Sb surface layer to show a strong $p_y - d_{z^2}$ orbital hybrid character in the occupied QW state.

Figure 4a shows the distribution of band feature from surface Sb(110) structure extending over the Au(111) SBZ with band feature around the high symmetry points of the Sb(110) SBZ symbolized by $\Gamma$, X, Y and S, respectively. Due to the three different azimuthal orientations of the Sb(110) rhombohedral lattice and the distinct lattice parameters between the Au(111) surface and the Sb(110) unit cell, the band features $\Gamma$ and X meet at the $\overline{K}$ point of the Au(111) SBZ, while the feature $\Gamma$ from the 2nd SBZ of the Sb(110) sub-unit cell appears independently at the $\overline{M}$ symmetry point of the Au(111) SBZ. Figure 4b and 4c show the experimentally recorded band dispersion along $\overline{\Gamma} - \overline{X}$ in the first Sb(110) SBZ (Figure 4b) and the second SBZ (Figure 4c) marked as blue lines in Figure 4a. We found that the Rashba-type splitting band $S_1$, as we discussed above, repeats periodically along $\overline{\Gamma} - \overline{X}$ with respect to the Sb(110) SBZ, since the symmetry path $\overline{\Gamma} - \overline{X}$ of the second Sb(110) SBZ corresponds to the path $\overline{M} - \overline{K}$ of the Au(111) SBZ. This Rashba-type splitting band $S_1$ shows a clear band dispersion along $\overline{\Gamma} - \overline{X}$ in the 2nd SBZ of Sb(110) (or $\overline{M} - \overline{K}$ of the Au(111) SBZ), however, it is partially shielded by the Au bulk state in the 1st SBZ of Sb(110) and highlighted by the red dashed curve in Figure 4b.

Interestingly, in Figure 4(c), it seems the feature of the Rashba-type splitting band connecting between $\overline{\Gamma}$ and $\overline{X}$ points shows a mirror symmetry with a mirrored plane positioned at the midpoint and perpendicular to the $\overline{\Gamma} - \overline{X}$ Brillouin path. This mirror-symmetry Rashba-type splitting band shows a stronger intensity of the band crossing around the $\overline{X}$ points than the crossing at $\overline{\Gamma}$ points, due to the signal from the $\overline{\Gamma}$ symmetry point meeting with the signal from the anther two $\overline{X}$ symmetry points in the 2nd Sb(110) SBZ (shown in Figure 4a). Carefully checking the energy of the crossing point at $\overline{X}$ point and $\overline{\Gamma}$ point in Figure 4c, it can be found that the binding energy of the Rashba splitting band at $\overline{\Gamma}$ point is slightly higher than at $\overline{X}$ point, which is in agreement with the energy separation of the two pairs of Rashba-type bands appearing at $\overline{K}$ point of the Au(111) SBZ, shown in Figures 3e-f.

As we claimed mirror symmetry in the Sb stripe structure perpendicular to $x$ axis and the nice agreement between the DFT calculated band structure and ARPES data, we do a further analysis of the role of a pair of antiparallel in-plane potential gradients along the $y$ direction plays on the band structure. The absence of mirror symmetry perpendicular to the $y$-axis engenders a potential gradient along the y-direction. This gradient, in turn, serves as the driving force behind the possible emergence of QW states originating from the Sb $p_y$ orbitals. Figure 4d shows the band structure unfolded to the SBZ of the Sb(110) sub-unit cell. The three pairs of QW states, marked by the black arrows in Figure 4d, can be clearly observed at $\overline{X}$ point and $\overline{\Gamma}$ point, respectively. The projected $p_x$ and $p_y$ states along the $\overline{Y} - \overline{\Gamma} - \overline{X} - \overline{Y}$ direction, overall, show a mirror symmetry with the mirror plane marked by the red dashed line in Figure 4d, which indicates the chiral properties of the three pairs of QW states at $\overline{X}$ point and $\overline{\Gamma}$ point. Hence, the mirrored character of the occupied Rashba-type band $S_1$ observed in our ARPES measurement and the calculated band structure projected from the Sb topmost layer indicates the chiral properties of the Rashba-type band $S_1$.

Contrary to the reported increasing spin splitting of QW states with the level of their binding energy in Bi(110) monolayer on Cu(111) system[11], the QW states of Sb monolayer on Au(111) surface show an inverse tendency of the Rashba spin splitting. Since Sb shows intrinsically weak SOC while SOC in

Au is strong, the degree of the Rashba spin splitting in different levels of QW states can be modified by hybridization with Au states. Besides the character of the Sb $p_y$ orbital appearing in the QW states, orbital decomposition of calculated band structure projected from the topmost Au layer indicates the contribution from the Au orbital. As we can see in Figure 4e, the occupied QW state at -1.60 eV shows a strong hybrid character between the Sb $p_y$ orbital (weighted by red on the left side of the panel) and the Au $d_{z^2}$ orbital (weighted by blue on the right side of the panel), the unoccupied QW state at 1.39 eV shows an Sb $p_y$-Au $p_x$ hybrid orbital character while the QW state at 2.37 eV shows a weak hybrid character with Au states(Figure S4b-c). Even though the higher QW state shows the tendency of increasing delocalization towards the vacuum due to the effective lowering of the vacuum barrier for the higher-lying states, we deduce that SOC plays a more important role in the Rashba spin splitting degree.

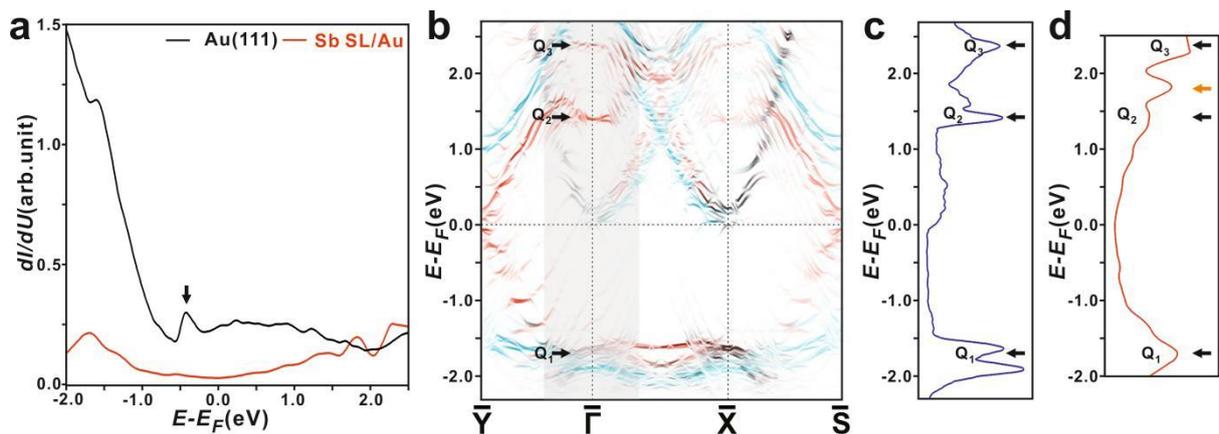

Figure 5. Electronic structures of QW states. (a) Typical $dI/dV$ spectrum obtained at the centre regions of a large terrace of 2D Sb stripe structure covered Au(111) (in red) and clean Au(111) (in black), acquisition conditions: U = +0.1 V, I = 0.9 nA, T = 4.0 K, B = 0 T. (b) Orbital decomposed band structure of the 2D Sb covered on the Au(111). The blue and red indicate, respectively, the $p_x$ and $p_y$ orbital contributions. (c) The local (density of states) DOS integrated over the zone-center region near the $\overline{\Gamma}$ point, marked by a gray shadow in (b), with a Gaussian broadening of 5.0 meV, in comparison with (d) our experimental $dI/dV$ spectrum (in red) of the 2D Sb film. The occupied and unoccupied QW states at $\overline{\Gamma}$ corresponding to the peak positions in (c) and (d) are marked with black arrows.

To determine the QW states induced by the absence of mirror symmetry perpendicular to the $y$-axis in 2D Sb stripe structure, tunneling spectra has been acquired in comparison with first-principle electronic structure calculations. Figure 5a displays the $dI/dV$ spectrum taken in a bias range from 2.5 to −2.0 V on both 2D Sb covered Au(111) (in red) and clean Au(111) (in black). On 2D Sb covered Au(111), the Shockley surface state at -0.43 V and an increasing background of density at high binding energy originating from clean Au(111) are obviously inhibited and several characteristic features appear, indicating the corresponding electronic states mainly originate from the topmost Sb adlayer. Based on the orbital-decomposed band structure of the 2D Sb stripe surface from the DFT calculations (Figure 5b), we further integrate the DOS around the $\overline{\Gamma}$ point as shown in Figure 5c. Note that Our DFT calculations fairly support the peaks at different energies corresponding to the QW states, marked by black arrows, observed in our STS measurements (Figure 5d), except for an additional peak at around 1.8 eV, originating from a hybrid state between Sb $p_z$ and Au $d_{z^2}$ orbitals (Figure S5).

**SUMMARY**

In our Letter, our STM observation confirmed the 2D Sb(110) rhombohedral crystal formed on Au(111). The Sb(110) unit cell possesses mirror symmetry with a pair of antiparallel potential gradients generated by the misalignment of the central Sb atom in two adjacent Sb(110) unit cells. we disentangle the intrinsic Sb(110) band from the Umklapp scattered branches of the Au $sp$ band and the three different azimuthal orientations of the (110) rhombohedral phase. Further, we studied the role of the symmetry breaking in the Sb(110) electronic structure. Our finding of three pairs of

QW-type Rashba Soc bands appearing at the $\overline{\Gamma}$ and $\overline{X}$ points with a chiral character is fully confirmed by ARPES measurements, STS spectra, and first-principles electronic structure calculations. We also explained the important role of the strong spin-orbital coupling within the Au states, which hybridizes with the Sb $p_y$ states to modify the spin splitting of the QW states.

**ACKNOWLEDGMENT**

This work was partially supported by the Research Council of Norway through its Centres of Excellence funding scheme, Project No. 262633, "QuSpin". Lina Liu and Yong Chen acknowledge the support from Villim Investigator program (Grant No. 25931). X. S. W. acknowledges support from the Natural Science Foundation of China (NSFC) Grant No. 12174093 and the Fundamental Research Funds for the Central Universities.

**Supporting information**

# Chiral Quantum well Rashba splitting in Sb monolayer on Au(111)

Jinbang Hu[1,+], Lina Liu[2], Xiansi Wang[3], Yong P. Chen[2], Justin W Wells[1,4,+]

*[1]Department of Physics, NTNU, Trondheim, Norway.*
*[2]Institute of Physics and Astronomy and Villum Centers for Hybrid Quantum Materials and Devices, Aarhus University, 8000 Aarhus-C, Denmark.*
*[3]Hunan University, Changsha, 410082, China.*
*[4]Semiconductor Physics, Department of Physics, University of Oslo (UiO), NO-0371 Oslo, Norway.*

[+]Corresponding author: jinbang.hu@ntnu.no.

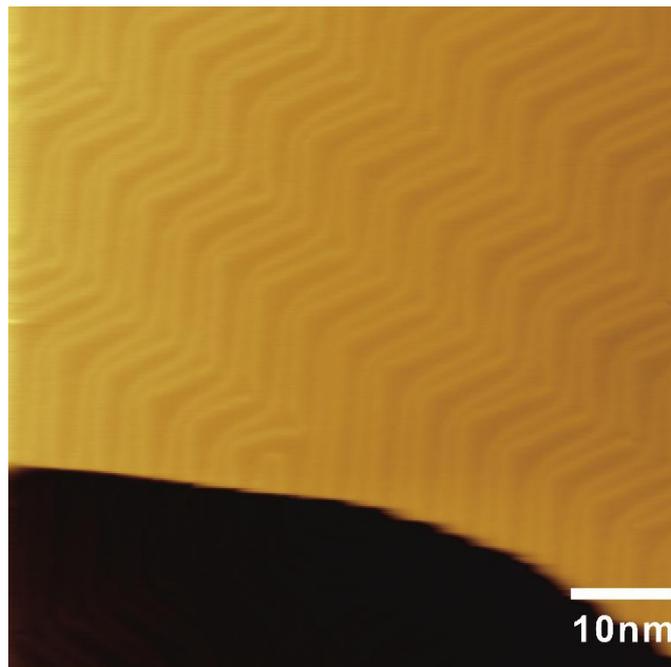

Figure S1. STM image of the clean Au(111) surface before Sb deposition. Scanning parameters: -1.2 V, 0.1 nA.

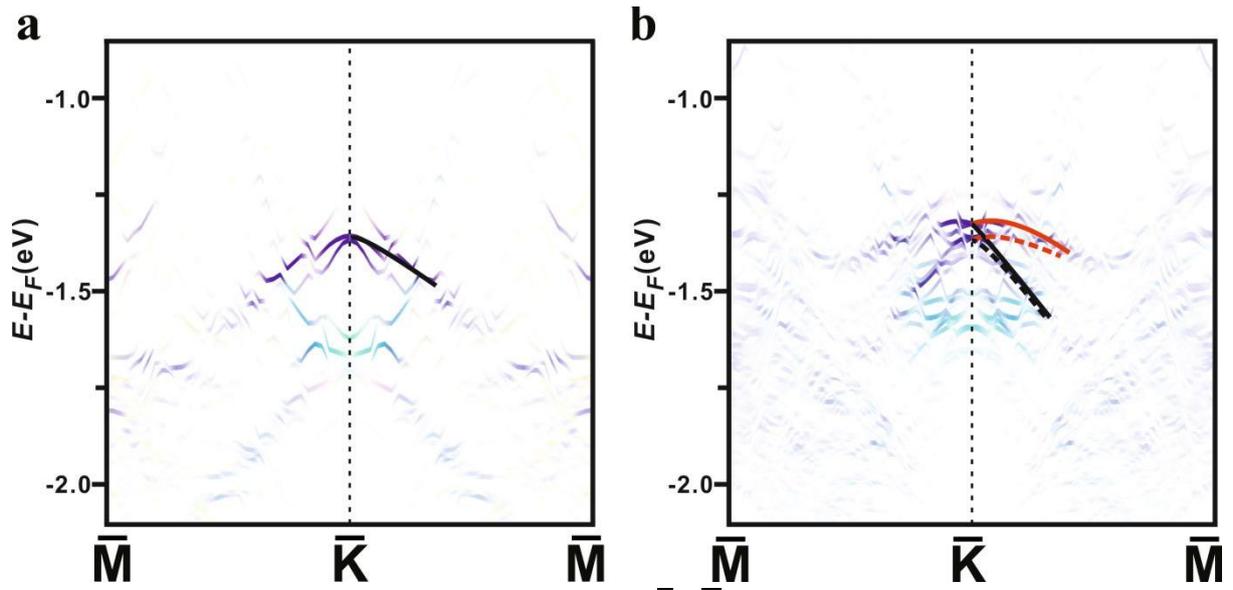

Figure S2. Comparison between calculated band dispersion along $\overline{K}-\overline{M}$ direction without SOC (a) and with SOC (b) around $\overline{K}$ point to illustrate the splitting of two pairs of Rashba-type band using a model of the 2D Sb(110) rhombohedral structure without buckling of Sb topmost layer. Blue and pink symbols show contributions from the $p_x$ and $p_y$ states, respectively.

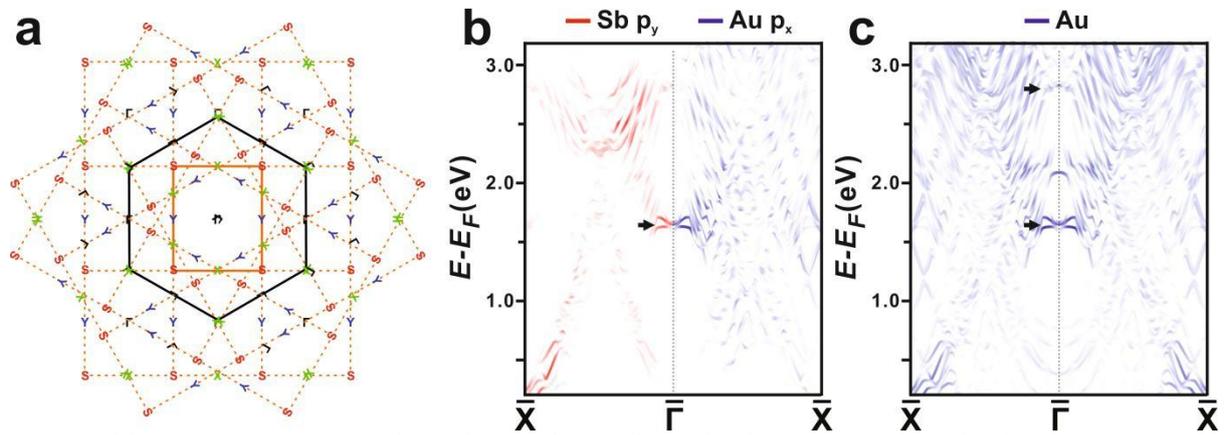

Figure S3. (a) A complete distribution of band features from surface Sb(110) structure around Au(111) SBZ due to the three different azimuthal orientations of the 2D Sb(110) rhombohedral lattices. The band structures around the high symmetry points of the Sb(110) SBZ are symbolled by Γ, X, Y, and S, respectively. (b) Calculated band structure projected from the $p_x$ orbital (weighted by blue in the right side of the panel) of topmost Au layer and the $p_y$ orbital (weighted by red in the left side of the panel) of topmost Sb surface layer shows a obvious Sb $p_y$-Au $p_x$ orbital hybrid character in the unoccupied QW state at 1.36 eV, marked by the black arrow. (c) Calculated band structure projected from the Au topmost layer shows a weak (strong) contribution from Au states to the QW state at 2.37 eV (1.36 eV), marked by the black arrow.

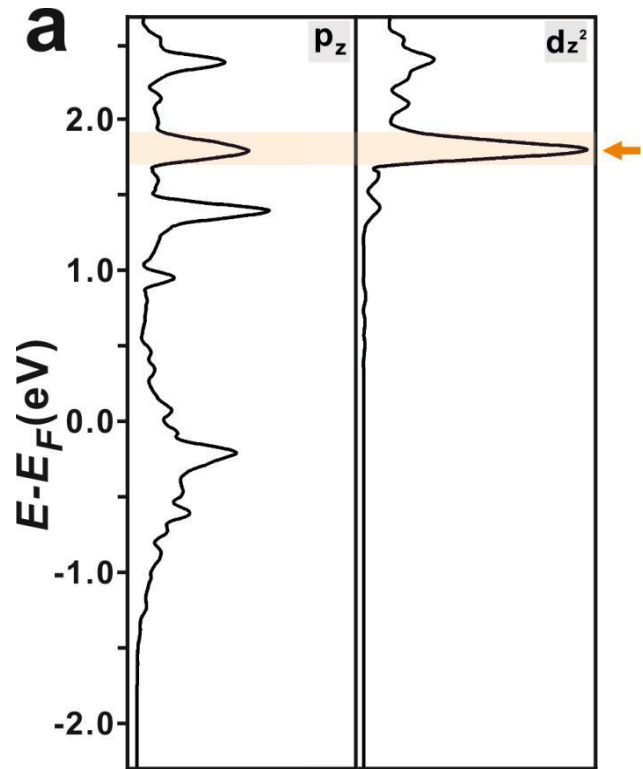

Figure S4. The Left (right) panel shows the partial DOS integrated over the zone-center region near the $\overline{\Gamma}$ point of Sb $p_z$ (Au $d_{z^2}$) orbital-decomposed band structure from the 2D Sb stripe surface (Au topmost layer).